\documentclass[twocolumn, aps]{revtex4}
\usepackage{amssymb}
\usepackage{bm}
\usepackage{psfig}

\newcommand{\rset}{{\mathbb{R}}}

\newcommand{\tset}{{\mathbb{T}}}

\newcommand{\tcset}{\tilde{\mathbb{C}}}
\newcommand{\tpartial}{\tilde\partial}
\newcommand{\tz}{\tilde{\bm z}}
\newcommand{\tnabla}{\tilde\nabla}
\newcommand{\tJ}{\tilde J}

\newif\iffigs
\figstrue % uncomment if figures are present
\iffigs
\fi
\def\drawing #1 #2 #3 {
\begin{center}
\setlength{\unitlength}{1mm}
\begin{picture}(#1,#2)(0,0)
\put(0,0){\framebox(#1,#2){#3}}
\end{picture}
\end{center} }

\def\ds{\displaystyle}

\def\v{{\bm v}}
\def\vr{\v_{\rm r}}
\def\vi{\v_{\rm i}}

\def\x{{\bm x}}

\def\y{{\bm y}}
\def\z{{\bm z}}
\def\k{{\bm k}}

\def\rf#1{(\ref{#1})}

\def\xfg#1{Fig.~\ref{#1}}

\def\un{\hbox{{1\kern -0.25em\raise 0.4ex\hbox{{\scriptsize $|$}}}}}

\def\e{{\rm e}}

\begin{document}
\title{Singularities of Euler flow? Not out of the blue!}
\author{U.~Frisch$^{\rm a,b}$,
        T.~Matsumoto$^{\rm a,b,c}$ and
        J.~Bec$^{\rm a,b}$}
\affiliation{%
$^{\rm a}$CNRS UMR 6529, Observatoire de la C\^ote d'Azur, BP 4229, 06304 Nice
          Cedex 4, France \\
$^{\rm b}$CNLS - Theoretical Division, LANL, Los Alamos, NM~87545,
          USA\\
$^{\rm c}$Dep. Physics, Kyoto University,
          Kitashirakawa Oiwakecho Sakyo-ku, Kyoto 606-8502, Japan}
%\draft
\date{\today}
\draft{\emph{J.\ Stat.\ Phys.}\/ in press. (updated version)\\}
\begin{abstract}
  Does three-dimensional incompressible Euler flow with smooth initial
  conditions develop a singularity with infinite vorticity after a
  finite time?  This blowup problem is still open. After
  briefly reviewing what is known and pointing out some of the
  difficulties, we propose to tackle this issue for the class of flows
  having analytic initial data for which hypothetical real
  singularities are preceded by singularities at complex locations.
  We present some results concerning the nature of complex space
  singularities in two dimensions and propose a new strategy for the
  numerical investigation of blowup.
 (A version of the paper with
  higher-quality figures is available at \hbox{http://www.obs-nice.fr/etc7/complex.pdf})
\end{abstract}
\vspace*{5mm}
\noindent{\em Dedicated to the memory of Richard Pelz.}
\maketitle

\section{Phenomenology of blowup}
\label{s:intro}
According to Richardson's ideas on high Reynolds number
three-dimensional turbulence, energy introduced at the scale $\ell_0$,
cascades down to a scale $\eta\ll \ell_0$ where it is dissipated.
Consider the total time $T_\star$ which is the sum of the eddy
turnover times associated with all the intermediate steps of the
cascade. From standard phenomenology \`a la Kolmogorov 1941 (K41), the
eddy turnover time varies as $\ell^{2/3}$.  If we let the viscosity
$\nu$, and thus $\eta$, tend to zero, $T_\star$ is the sum of an
infinite {\it convergent\/} geometric series.  Thus it takes a {\it
  finite time\/} for energy to cascade to infinitesimal scales, an
observation first made by Onsager \cite{onsager49}.  We also know that
in the limit $\nu\to 0$, the enstrophy, the mean square vorticity,
goes to infinity as $\nu^{-1}$ (to ensure a finite energy
dissipation).

From such observations, it is tempting to conjecture that ideal flow,
the solution of the (incompressible) 3-D Euler equation
\begin{eqnarray}
\partial_t{\bm v} +{\bm v}\cdot\nabla{\bm v} &=& -\nabla p, \\
\nabla\cdot{\bm v}&=&0,
\label{euler}
\end{eqnarray}
when initially
regular, will
spontaneously develop a singularity in a finite time. 

This is of course incorrect: the kind of phenomenology assumed above
is meant only to describe the (statistically) steady state in which
energy input and energy dissipation balance each other; the inviscid
($\nu=0$) initial-value problem is not within its scope. Another
possible argument in favor of singularities has to do with the scaling
properties of the high Reynolds number solutions (e.g.\ the $k^{-5/3}$
spectrum). For the simpler case of the Burgers turbulence \cite{houches81},
the scaling of
spectra and structure functions is clearly rooted in the singularities
(shocks) appearing in the solutions in the limit of vanishing
viscosity. It is however well-known that power-law behavior can be
present without any singularities. An example is the Holtsmark
process, that is any component of the electric or gravitational field
produced at a given point by a set of charges or masses with an
initial Poisson distribution in space and moving with uniform
independent isotropic velocities (having, e.g., a Gaussian
distribution). It is then easily shown (by adaptation of the technique
used by Chandrasekhar \cite{Chandrasekhar43}) that the correlation
function is $\propto |t-t'|^{-1}$. This power-law behavior comes from
the algebraic distribution of the distances of closest approach to the
point of measurement and not from singularities of individual
realizations (which are actually analytic).

There is yet another phenomenological argument, not requiring K41,
which suggests finite-time blowup of the vorticity.  Consider the
equation for the vorticity ${\bm \omega} \equiv \nabla \wedge{\bm v}$
for inviscid flow, written as
\begin{equation}
D_t{\bm \omega}={\bm \omega} \cdot \nabla {\bm v},
\label{lagvort}
\end{equation}
where $D_t\equiv \partial_t+{\bm v}\cdot\nabla$ denotes the Lagrangian
derivative. Observe that $\nabla {\bm v}$ has the same dimensions as
$\bm \omega$ and can be related to it by an operator involving
Poisson-type integrals. (For this use the fact that $\nabla^2{\bm v}=
-\nabla\wedge \bm \omega$.) It is then tempting to predict that the
solutions of (\ref{lagvort}) will behave as the solution of the scalar
nonlinear equation
\begin{equation}
D_t s=s^2,
\label{scalvort}
\end{equation}
which blows up in a time $1/s(0)$ when $s(0)>0$.  Actually,
(\ref{scalvort}) is just the sort of equation one obtains in trying to
find rigorous {\it upper bounds\/} to various norms when studying the
well-posedness of the Euler problem.  This is precisely why the
well-posedness `in the large' (i.e.\ for arbitrary $t>0$) is an open
problem in three dimensions.

\begin{figure}[ht]
\iffigs
\centerline{\psfig{file=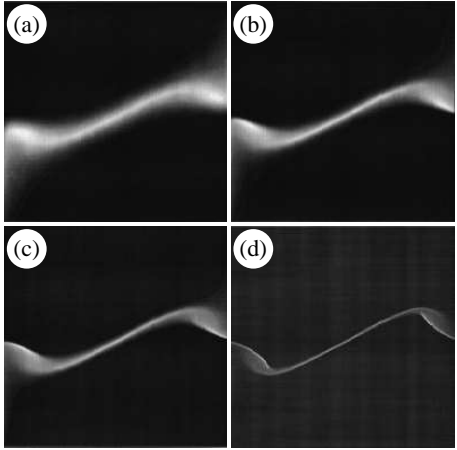,width=6cm,clip=}} 
\else\drawing 60 10 {vorticity pancake}
\fi
\caption{Simulation of the formation of a vorticity pancake at four different
times from Ref. \cite{brachetetal92}}
\label{f:pancake}
\end{figure}

The evidence is that the solutions of the Euler equation behave in a
way much tamer than predicted by (\ref{scalvort}) because of a phenomenon
known as ``depletion of nonlinearity'': when small-scale structures appear 
through the nonlinear dynamical evolution from smooth initial data, they tend
to display, at least locally, a much faster dependence on one particular 
spatial direction, so that the flow is to
leading order one-dimensional \cite{frischetal83}. An example in three
dimensions are the vorticity pancakes seen in simulations, as illustrated in 
Figure~\ref{f:pancake}. If the flow were exactly
one-dimensional, the nonlinearity would vanish (as a consequence of
the incompressibility condition). 

Depending on how strong this depletion is and also on how persistent
it is, finite-time blowup may or may not occur \cite{Frisch95,constantin-btl}. 

A review of the known mathematical results concerning the initial
value problem for the Euler equation may be found in Ref.\ 
\cite{majda-bertozzi}. We just mention a few salient facts.  In two
dimensions, for sufficiently smooth initial data in a bounded domain
(including periodic boundary conditions) or with sufficiently fast
decay at large distances, H\"older continuity of the vorticity is
preserved for all times. The key result was actually obtained in 1933
\cite{wolibner,holder}. In three dimensions, for sufficiently smooth
initial data, regularity is guaranteed only for a finite time. The
first such result goes back to 1925 \cite{lichtenstein}. A very
important result, established in the late eighties by Beale, Kato and
Majda (BKM) is that blowup, if it takes place, requires the
time-integral of the supremum of the vorticity, and hence the
vorticity itself, to become infinite \cite{bkm}. As pointed out in
Ref.\ \cite{majda-bertozzi}, the main stumbling block in trying to improve
the existing 3-D regularity results beyond a finite time is our still
very rudimentary understanding of the mathematics of depletion. It is
worth stressing here that an even partial progress on depletion could
play a crucial role in understanding regularity issues for the 3-D
Navier--Stokes equation \cite{claymath}.

Numerical studies of 3-D blowup have been going on since at least the
seventies with the development of spectral methods
\cite{orszag-les-houches77}. The advantage of spectral methods is that
they allow the kind of very high accuracy which is desirable in
investigating possible singular behavior. When spectral methods were
able to achieve fairly high resolutions ($256^3$ or more grid points),
evidence for depletion emerged, possibly of sufficient strength to
prevent blowup \cite{brachetetal} (see also Section~\ref{s:tracing}).
It was however realized that singularities (or near singularities) of
the 3-D Euler are highly localized in space and that the kind of
uniform grid used in standard spectral methods is very wasteful. A
number of new methods were developed using nonuniform, e.g., adaptive
meshing and often rather special initial conditions. Such studies have
provided some evidence for blowup, based on extrapolating the behavior
in time of the supremum of the vorticity. For the convenience of the
reader we give a list of the key references (provided to us by R.~Pelz
and R.~Kerr). It is not our purpose here to review such work.

We shall argue in this paper that there exists a class of analytic
flows for which numerical studies of blowup can probably be carried
out with sufficient control to distinguish genuine and spurious
numerical blowup.  In Section~\ref{s:analytic}, we recall some known
facts about the Euler equation in the complex domain. In
Section~\ref{s:tracing}, we first recall how to detect precursors of
blowup by tracing complex space singularities with a spectral method;
then we propose a new spectral adaptive approach capable, in principle,
to resolve highly localized singularities. For this method it can be
crucial to have information about the nature of complex space
singularities.  Sections~\ref{s:complex2D} and~\ref{s:asymptotics}
report preliminary results concerning the singularities of the Euler
equation in two space dimensions. Section~\ref{s:conclusion} presents
concluding remarks.

\section{Analyticity of  solutions to the Euler equation}
\label{s:analytic}

We are here interested in solution of the $d$-dimensional Euler
equation \rf{euler} with real analytic initial data $\v_0$, extended
into the complex domain, a question addressed also in 
Refs.~\cite{tanveer-speziale,BNXW}.  For simplicity, we assume ``periodic
boundary conditions'', that is space periodicity with period 1 in all
$d$ coordinates (although we use $2\pi$ when discussing numerical
results). The configuration space is thus $\tcset^d \equiv \tset ^d +i
\rset^d$, where $\tset^d$ is the $d$-dimensional periodicity torus.
The analytic continuation of the solution to $\tcset^d $ is denoted
$\v(\z,t)$, where $\z\equiv \x+i\y$  (we do not complexify the time
variable).

We now give a brief survey of some key results for the Euler equation in the
complex domain. Since this paper is intended mostly for a readership 
of fluid dynamicists, physicists and numerical analysts, we shall avoid
using excessively formal mathematical language but, of course, distinguish
clearly what is truly proven from what is just conjectured.

The first results about analyticity of the solution to the Euler
equation have been obtained, to the best of our knowledge, in the
seventies.  \emph{With periodic boundary conditions, analyticity
  assumed initially, is preserved for all time in two dimensions}\/
\cite{bardos-benachour-zerner} \emph{and at least for a finite time in
  three dimensions}\/ \cite{baouendi-goulaouic,benachour76}.
The simplest derivations of such results are obtained by using
  Lagrangian methods \cite{bardos-benachour,benachour78}. One
follows complex characteristics, the solutions of $\dot\z=\v(\z,t)$,
in order to control the width $\delta(t)$ of the ``analyticity
strip'', that is the distance from the real domain of the nearest
singularity in $\tcset^d $ for the solution at time $t$.

What singles out two dimensions is that vorticity is conserved along
the characteristics (as it is in the real domain). The key estimate
used to prove all-time analyticity is that when $\delta(t)>s$ and
$0<s<1$, one has
\begin{eqnarray}
&&|{\rm Im}\, \v(\x+i\y,t)| \le
 B_s||{\bm \omega}(\cdot,t)||_s\, |\y|\,\ln{1\over|\y|}, \label{supimv}\\
&& ||{\bm \omega}(\cdot,t)||_s \equiv \sup_{|\y|\le s,\, \x \in\tset^2} 
|{\bm \omega}(\x+i\y,t)|, 
\label{defomegas}
\end{eqnarray}
where $B_s \equiv C_1+C_2 \left(\ln (1/s)\right)^{-1}$ and $C_1$ and
$C_2$ are positive constants. This is a consequence of the
Biot--Savart law relating the velocity and the vorticity.  Actually,
\rf{supimv} is mostly a reworking in the complex domain of an estimate
obtained in Refs.~\cite{wolibner,holder} for proving all-time
regularity in the real domain.  It follows from \rf{supimv} and
vorticity conservation that, if initially $\delta(0)>s$, at any later
time the width of the analyticity strip has a double exponential lower
bound $\delta(t)> s^{\exp (B_s ||{\bm \omega}_0||_s t)}$. Hence
analyticity holds for all times, but complex singularities can get
arbitrarily close to the real domain.

In three dimensions, vorticity is not conserved since it can be
stretched by velocity gradients. As a consequence, very poor control
of $\delta$ is available and its vanishing after a finite time cannot
be ruled out.  Still, it is easy (but a bit technical) to show that,
if at some time $t_0$ the solution is analytic, it will stay so in a
(possibly small) time interval $[t_0,t_0+T_\star[$. Furthermore it was
shown by Benachour and Bardos that, if $\delta(t_0)>s$ then, in this
time interval, one has the square root lower bound $\delta(t)>
(s/2)\left(1-(t-t_0)/T_\star\right)^{1/2}$
\cite{benachour76,benachour78,bardos-benachour}. One important
consequence, is that: \emph{In three dimensions with periodic boundary
  conditions and analytic initial conditions, analyticity is preserved
  as long as the velocity is continuously differentiable}\/ ($C^1$)
\emph{in the real domain}\/ \cite{bardos-benachour}. The BKM theorem
(cf.\ Section~\ref{s:intro}) allows us to strengthening this result:
analyticity is actually preserved as long as the vorticity is finite.

As long as $\delta>0$ the Euler equation can be written not only in the real
domain but directly in the complex domain. In particular we shall find it
useful to write it on ``parareal domains''. By this we understand a domain of
the form $\tset^d+i\y$ for fixed $\y$ such that $|\y|<\delta$, i.e.\ obtained
from the real (periodic) domain $\tset^d$ by a fixed imaginary shift. In such
a domain the velocity field is complex.  By separating the velocity and the
pressure into real and imaginary parts, $\v =\vr +i\vi$ and $p=p_{\rm
r}+ip_{\rm i}$, we can rewrite the Euler equation on a parareal domain as
\begin{eqnarray} 
&&\partial_t{\vr} +{\vr}\cdot\nabla{\vr} - {\vi}\cdot\nabla{\vi}= -\nabla
p_{\rm r},\label{eulerr}  \\
&&\partial_t{\vi} +{\vr}\cdot\nabla{\vi} + {\vi}\cdot\nabla{\vr}= -\nabla
p_{\rm i}, \label{euleri} \\
&&\nabla\cdot{\vr}=0, \quad \nabla\cdot{\vi}=0.
\label{eulerparadiv}
\end{eqnarray}
Note that eqs.~(\ref{eulerr}-\ref{eulerparadiv}) have some similarity
to the magnetohydrodynamics (MHD) equations, with $\vr$ and $\vi$
playing the role of the velocity field and the magnetic field,
respectively.  The main differences are the following: (i) the
presence of a pressure term in the second equation (there is no such
term in the MHD induction equation); (ii) if we linearize the
equation around a uniform $\vi$, we do not obtain Alfv\'en-like waves
but an instability whose growth rate is proportional to the
wavenumber. The interpretation of this instability is that the
imaginary translation induced by an imaginary velocity corresponds to
an exponential factor in Fourier space (rather than a phase factor
which would be associated to a real translation). We also observe that
a real-imaginary decomposition similar to (\ref{eulerr})-(\ref{eulerparadiv})
has been used in Ref.~\cite{grujic-kukavica} for the Navier--Stokes equation
in the complex domain (in order to estimate the space analyticity radius
of solutions in terms of $L^p$ and $L^\infty$ norms of initial data).

The proof of the local-in-time analyticity result of
Ref.~\cite{baouendi-goulaouic} is easily extended to the Euler
equation in a parareal domain with analytic initial data in both two
and three dimensions. All-time analyticity is now ruled out, even in
two dimensions, since complex space singularities will generally cross
after a finite time any parareal domain as they approach the real domain.

A consequence of local-in-time analyticity is that \emph{the width of
the analyticity strip $\delta(t)$ cannot decrease discontinuously in
time.}\/ This is a consequence of the Benachour--Bardos square root
lower bound on $\delta(t)$ given above. In particular if there is
finite-time blowup, that is $\delta(t)$ vanishes at some $t_\star$,
the vanishing takes place continuously: \emph{real singularities do not
come out of the blue}\/ (but out of the complex). This observation
has led to the introduction of the method of tracing complex space
singularities discussed in the next section.

Note that similar results about finite-time analyticity and continuity of
$\delta(t)$ can be obtained without periodic boundary conditions, provided the
solutions decrease sufficiently fast at large distances
\cite{benachour78}. However, if the solutions are not well behaved at infinity
the results may be wrong.  A counterexample, due to S.~Childress and E.~Spiegel
(quoted in \cite{rose-sulem}), is given by
\begin{eqnarray}
&&\!\!\!\!\!\!\!\!\v(x_1,x_2,x_3,t) = \left( {x_2+x_3\over t-t_\star},{x_3+x_1\over
t-t_\star}, {x_1+x_2\over t-t_\star}\right),\label{childspiegv}\\
&& \!\!\!\!\!\!\!\!p(x_1,x_2,x_3,t) = -{x_1^2+x_2^2+x_3^2\over (t-t_\star)^2}.
\label{childspiegp}
\end{eqnarray}

\section{Tracing complex singularities}
\label{s:tracing}

As suggested in Refs.~\cite{tracing,houches81}, when the initial data
for the Euler equation are analytic (and periodic), it is possible to
trace the temporal behavior of the width of the analyticity strip
$\delta(t)$ in order to obtain evidence for or against blowup. This
takes advantage of the signature of complex space singularities in
Fourier space.  In one dimension, it is well known that an analytic
function having isolated singularities in the complex plane has a
Fourier transform whose modulus decreases at large wavenumbers $k$ as
$\exp(-\delta k)$ (up to algebraic prefactors), where $ \delta$ is the
distance from the real domain of the nearest complex space singularity
\cite{carrier-krook-pearson,frisch-morf}.

So far, nothing was known concerning the nature of complex space
singularities of the multi-dimensional Euler equation. In more than
one dimension, singularities in the complex domain are never
point-like. For example, it may be shown that for the $d$-dimensional
(potential) inviscid Burgers equation with analytic initial data, the
singularities in the complex domain (before shock formation) are on
$(d-1)$-dimensional complex manifolds, a result which cannot be ruled
out for the Euler equation.  In Fourier space, consider wavevectors of
the form $\k =k \hat \k$ where $\hat \k$ is unit vector of fixed
direction. For fixed $t$, it may then be shown that, as $k\to\infty$,
the modulus of the Fourier transform decreases as
$\exp(-\delta(\hat\k,t) k)$ where $\delta$ now depends on the direction.
The width of the analyticity strip at time $t$ is then the minimum of
$\delta(\hat\k,t)$ over all directions.

\begin{figure}[ht]
\iffigs
\centerline{\psfig{file=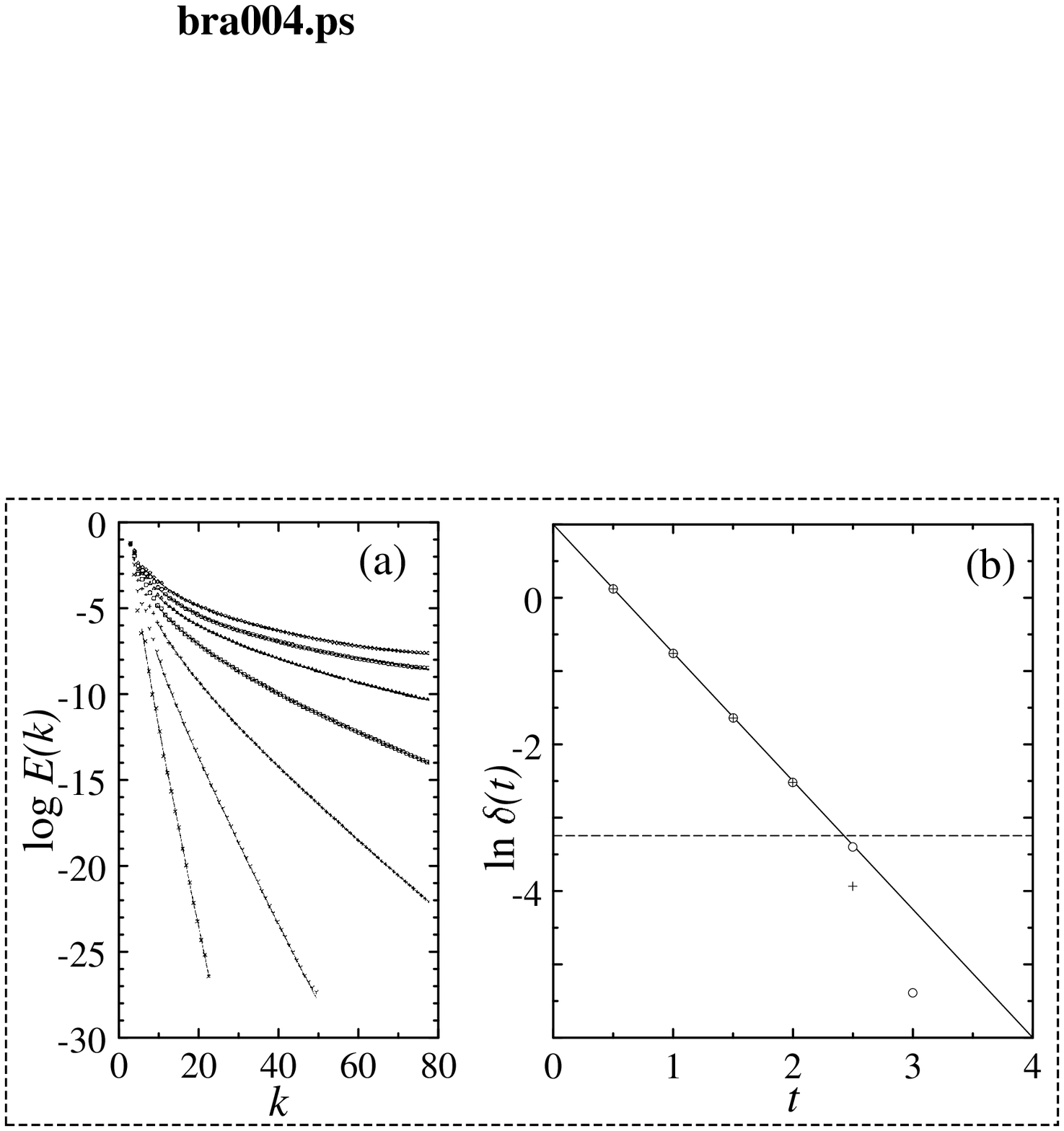,width=.48\textwidth,clip=}} 
\else\drawing 60 10 {\xfg{f:deltaTG} Taylor Green spectral}
\fi
\caption{Spectral simulation of the inviscid Taylor--Green vortex using
$256^3$ Fourier modes. (a) Evolution of the energy spectrum in lin--log
coordinates; from bottom to top: output from time $t=0.5$ in increments of
$0.5$. (b) Time-dependence of the width of the analyticity strip $\delta(t)$
in linear--log coordinates; the circles and plus signs correspond to $256^3$
and $128^3$ Fourier modes, respectively; the dashed line gives the threshold
of reliability (from Ref.~\cite{brachetetal}).}
\label{f:deltaTG}
\end{figure}

\begin{figure}[ht]
\iffigs
\centerline{\psfig{file=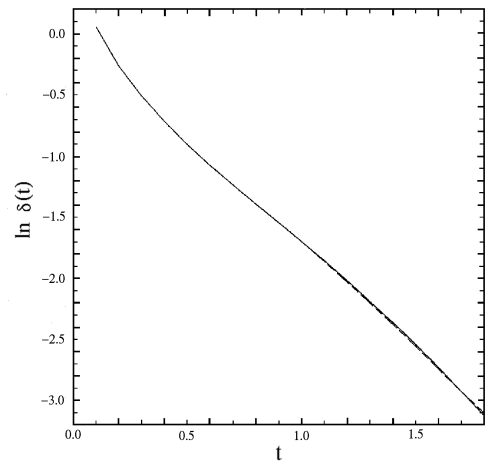,width=.37\textwidth,clip=}}
\else\drawing 60 10 {\xfg{f:deltanonsym} delta(t)}
\fi
\caption{Width of the analyticity strip
$\delta(t)$ from a spectral simulation of a periodic flow with no particular
symmetry. Dashed line data are for $200^3$ modes and 
solid line data for $256^3$ modes (from Ref.~\cite{brachetetal92}). } 
\label{f:deltanonsym}
\end{figure}

The numerical tracing of complex singularities is easy if the Euler
equation is integrated by a (pseudo-)spectral method
\cite{gottlieb-orszag} with enough spatial resolution to capture the
exponential tails in the Fourier transforms. This method was applied
for the first time to the three-dimensional Euler flow generated by
the Taylor--Green initial conditions \cite{brachetetal}
\begin{equation}
\left.
\begin{array}{rl}
\ds v_1\!\!\!\!&\ds =\,\sin x_1\cos x_2 \cos x_3,\\[1.6ex]
\ds v_2\!\!\!\!&=\ds \,-\cos x_1 \sin x_2 \cos x_3,\\[1.6ex]
\ds v_3\!\!\!\!&=\ds
\,0.
\end{array}
\right\}
\label{TaylorGreen}
\end{equation}
The simplest is then to plot  for various
times the ``energy spectrum'' $E(k,t)$, that is the angle-averaged
squared modulus of the Fourier transform of the velocity. By a
steepest descent argument one has (up to algebraic prefactors) $E(k,t)
\propto \exp (-\delta(t)k)$.  This is illustrated in
Figure~\ref{f:deltaTG}, taken from Ref.~\cite{brachetetal}, giving
for the first time evidence that the Taylor--Green vortex may actually
not have any blowup since $\delta(t)$ appears to decrease
exponentially in time. This behavior is observed reliably over an
interval of time during which $\delta(t)$ decreases by about one
decade. Later work, extending the simulations from a $256^3$ to a
$864^3$ grid, have confirmed this behavior over a range of one and a
half decade in $\delta$ (see Figure~\ref{f:deltanonsym} taken from 
Ref.~\cite{brachetetal92}).

The Taylor--Green vortex may just happen not to be a good candidate
for blowup. It is special in at least two ways. First, it has a lot of
symmetry 
(used to simplify the computation). Second, its vortex lines
have non-generic topology: they are closed instead of displaying the
KAM disorder that is typical of the integral lines of
three-dimensional divergenceless vector fields. General periodic flow
not displaying such pathologies has also been investigated in
Ref.~\cite{brachetetal92}.  It seems to give exponential shrinking of
$\delta$, but over just about half a decade in $\delta$. Two causes
for this reduced range are clear: the limited resolution permitted ten
years ago for flow without symmetry ($256^3$) and a crossover
phenomenon between two different small-scale structures localized at
different spatial locations, resulting into two different regimes with
a changeover around $t=1.05$ (cf. Figure~\ref{f:deltanonsym}).

If we could simulate general periodic flow in such a way as to observe
the variation of $\delta$ over several decades we would probably be able
to get good evidence for or against blowup. One way to achieve
this is just ``patience''. We can indeed expect to gain a little more than 
a factor four in 
spatial resolution every ten years: this requires  an increase in CPU
power of $4^4=256$, made possible by Moore's law. Another way is to 
renounce spectral methods and switch to adaptive methods. Unfortunately,
such methods had so far only finite-order accuracy and could not be used
to analytically continue the solutions into the complex domain, so that
we must renounce  measuring $\delta(t)$.

\begin{figure}[ht]
\iffigs
\centerline{\psfig{file=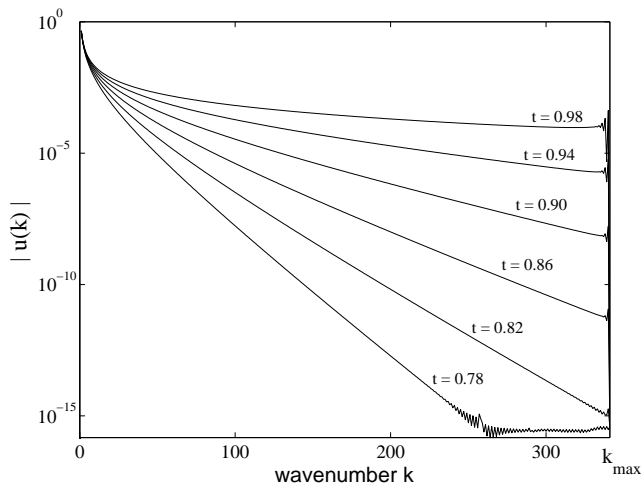,width=.48\textwidth}} 
\else\drawing 60 10 {\xfg{f:Tb} Tb}
\fi
\caption{Evolution of the modulus of the
Fourier transform for the solution of the one-dimensional Burgers equation
with initial condition $u_0(x)=\sin x$, for which the first real singularity
is at $t=1$. Around $t=0.82$ round-off noise disappears from the resolved
spectral range. A rather moderate resolution of $1024$ modes is used in
order to make the round-off noise fluctuations more visible.} 
\label{f:Tb}
\end{figure}

We propose here a new strategy, the ``spectral adaptive'' method which
combines the highly localized refinement permitted by adaptive methods with
full contact to the complex space structure.  This method, which is
still in the testing phase, will only be briefly outlined here. The basic idea
is to run a standard spectral simulation until the latest time $t_B$ when
$\delta(t)$ can be measured with very high accuracy and then to perform a
``regularizing analytic transformation'' $B$ on $\tcset^3$ with the following
properties: (i) it preserves $\tset^3$ globally and thus preserves
periodicity, (ii) it maps the complex singularities of the solution at time
$t_B$ away from $\rset^3$ (possibly to complex infinity). In the new
coordinates resulting from the transformation $B$, the problem is still
periodic and can again be integrated by a suitable spectral method (with new
difficulties since the coefficients are now strongly non-uniform). The
procedure can, in principle, be repeated several times. The spectral adaptive
strategy is yet to be fully implemented. So far, we have performed tests
in one space dimension on the Burgers equation. These have revealed
that, in order to minimize errors, it is best to perform the regularizing
transformation around the time when the round-off noise just disappears
from the tail of the spatial Fourier transform of the solution, which is a
function of the resolution and of the round-off level (cf. Figure~\ref{f:Tb}).

For one-dimensional equations, complex
singularities are point-like and we know simple transformations
which have the required properties. In higher dimensions,
singularities are on extended objects and their nature is not well
known. 
A possible candidate for the regularizing analytic
transformation $B$ could be  the \emph{inverse Lagrangian map}\/ between time
$t_{B}$ and the initial time (also called back-to-labels
map \cite{constantin-btl}). This is definitely the case for the 
Burgers equation in any dimension, whose solution stays entire in 
Lagrangian coordinates if the initial condition is entire, that is
analytic in the whole complex domain. However 
recent results for the two-dimensional incompressible Euler equation, the 
details of which will be published elsewhere, indicate that entire 
initial conditions develop complex-space singularities  in both 
Eulerian and Lagrangian coordinates.

Note that the two-dimensional case is not just an
academic problem, as one might infer incorrectly from the proven fact
that 2-D Euler flow never blows up. One reason is that there is
evidence that 2-D Euler flow is much tamer than predicted by the
double exponential lower bound for $\delta(t)$ given in
Section~\ref{s:analytic}. Indeed, spectral simulations indicate that
$\delta(t)$ is actually decreasing exponentially \cite{tracing} and
that there is strongly depleted nonlinearity. Another reason is the
existence of two-dimensional variants of the Euler equation for which
blowup has still not been ruled out, such as axisymmetric flow with
swirl or 2-D free convection
\cite{GrauerSideris,PumirSiggia_swirl,EShu,CaflischErcolaniSteele}.

To find suitable candidates for the analytic transformation $B$ in two
dimensions we would like to know something about the complex singularities
of two-dimensional Euler flow with analytic initial conditions.
This is the subject of the next section.

\section{Numerical study of complex singularities for 2-D Euler flow}
\label{s:complex2D}

\begin{figure}[ht]
\iffigs
\centerline{\psfig{file=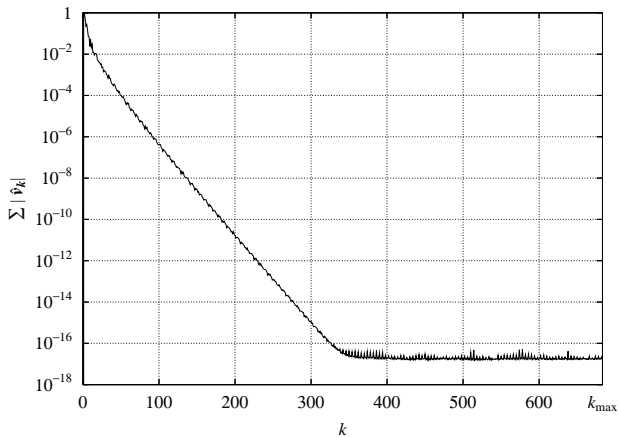,width=.48\textwidth}} 
\else\drawing 60 10 {\xfg{f:roundoff} roundoff noise shown}
\fi
\caption{Angle average, at a given time, of the modulus of the Fourier
transform of velocity in lin-log coordinates. The exponential
falloff reaches here the roundoff noise level for wavenumbers of
the order of 350, while the maximum wavenumber $k_{\max}$
determined by the $2/3$ dealiasing rule is 682 for a resolution
of $2048^2$.} 
\label{f:roundoff}
\end{figure}

Spectral simulations of 2D Euler flow with analytic initial conditions
give very strong evidence for the presence of complex space singularities
through the presence of exponential tails in the Fourier transforms of the
solution.
Can we find out something about the nature of such singularities?  
One way is to analytically continue the solution at time $t$ to complex 
locations, using the Fourier series. Suppose we have the Fourier
representation
\begin{equation}
\v(\x,t) =\sum_\k \e ^{i\k\cdot\x} \hat\v_\k(t),
\label{fourier}
\end{equation}
we can just substitute $\z=\x+i\y$ for $\x$ and obtain $\v(\z,t)$ as long as
the series converges, that is for $\y<\delta(t)$. Alternatively, we can
analytically continue the initial condition to the parareal domain
$\tset^2+i\y$ (a square in the 2D case) and then integrate the parareal
equations \rf{eulerr}-\rf{eulerparadiv} from $0$ to $t$.  Mathematically, the
two procedures are equivalent since they differ only by an imaginary
translation, that is an overall exponential factor $\e ^{-\k\cdot\y}$. Their
numerical implementations in high-resolution spectral simulations may however
not be equivalent, because of the presence of roundoff noise. At early times,
when $\delta(t)$ can be quite large, the exponential falloff of the Fourier
transform reaches roundoff noise level for wavenumbers which are much smaller
than the maximum wavenumber permitted by the simulation
(cf. Figure~\ref{f:roundoff}).
If we multiply this by the exponential factor $\e ^{-\k\cdot\y}$, roundoff
noise may be tremendously amplified (for $\k\cdot\y<0$). If we directly
integrate in the parareal domain we can, at each time step, identify and set
to zero the modes which are affected by roundoff, a procedure closely related
to that used by Krasny for integration of the vortex sheet problem
\cite{krasny}.  As $t$ increases, these modes will shift to higher
wavenumbers, eventually reaching the maximum wavenumber permitted by the
resolution. We are then effectively making better use of the available
resolution. We call this procedure parareal integration with noise
suppression.  This  method is rather crude and might be
improved by taking into account the strong anisotropy present near a
singularity.

We have used this method to integrate the two-dimensional parareal Euler
equation using from $512^2$ to $2048^2$ modes and initial conditions which are
trigonometric polynomials in the space coordinates and thus entire
functions. (As we shall see later, such initial conditions are also amenable
to a short-time expansion.) The cleanest results are obtained when the initial
condition $\v_0(\x)$ has only two modes
\begin{eqnarray}
&&\v_0(\x)= (\partial_2\psi_0(\x),\,-\partial_1\psi_0(\x)),
\label{defpsi0}\\
&& \psi_0(\x) = \cos (x_1) + \cos (2x_2),
\label{deuxmodes}
\end{eqnarray}
where $\psi(\x,t)$ is the stream function and $\partial_1$ and $\partial_2$
are the derivatives with respect to $x_1$ and $x_2$.

\begin{figure}[ht]
\iffigs
\centerline{\psfig{file=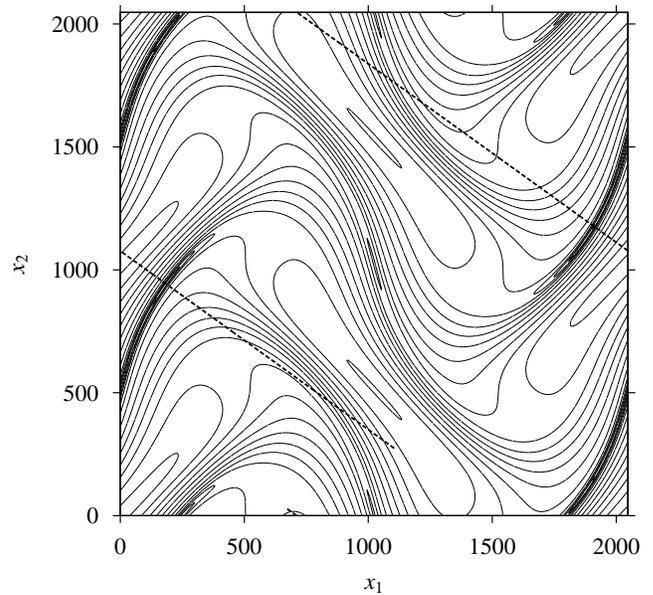,width=.48\textwidth,clip=}} 
\else\drawing 60 10 {\xfg{f:omega_cntr}snapshot real vort.}
\fi
\caption{Snapshot of the contours of the real part of vorticity in
the parareal domain $\bm y = (2\pi/64) (\cos \pi/6, -\sin \pi/6)$
just before the appearance of a singularity.} 
\label{f:omega_cntr}
\end{figure}

\begin{figure}[ht]
\iffigs
\centerline{\psfig{file=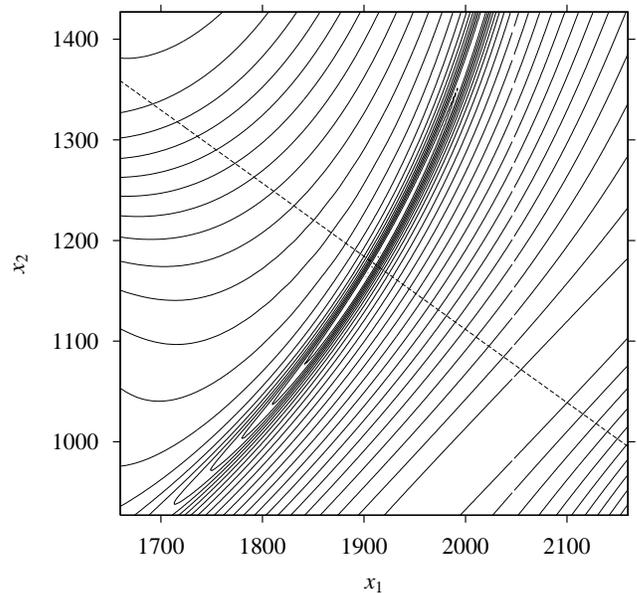,width=.48\textwidth,clip=}} 
\else\drawing 60 10 {\xfg{f:omega_cntr_enlar} enlargem. real vort.}
\fi
\caption{Enlargement of Figure~\ref{f:omega_cntr} showing the contour
lines of the real part of vorticity in the neighborhood of the
location where the singularity appears.} 
\label{f:omega_cntr_enlar}
\end{figure}

\begin{figure}[ht]
\iffigs
\centerline{\psfig{file=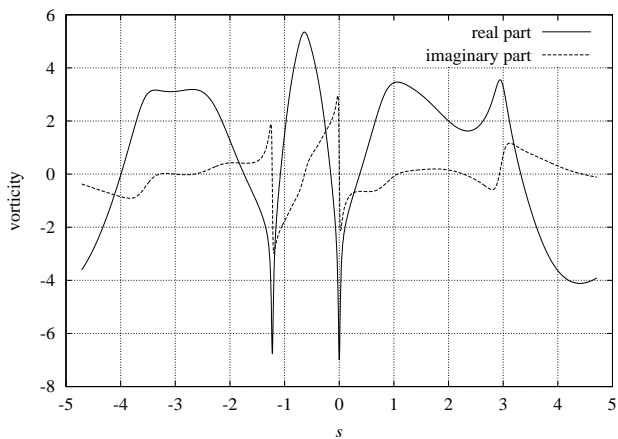,width=.48\textwidth,clip=}} 
\else\drawing 60 10 {\xfg{f:section_all} cut or real an imag.}
\fi
\caption{Real and imaginary parts of vorticity along the cut
represented as a dashed line on Figure~\ref{f:omega_cntr}. Both
real and imaginary parts of vorticity display quasi-singular
behavior near $s=-1.2$ and $s=0$.} 
\label{f:section_all}
\end{figure}

\begin{figure}[ht]
\iffigs
\centerline{\psfig{file=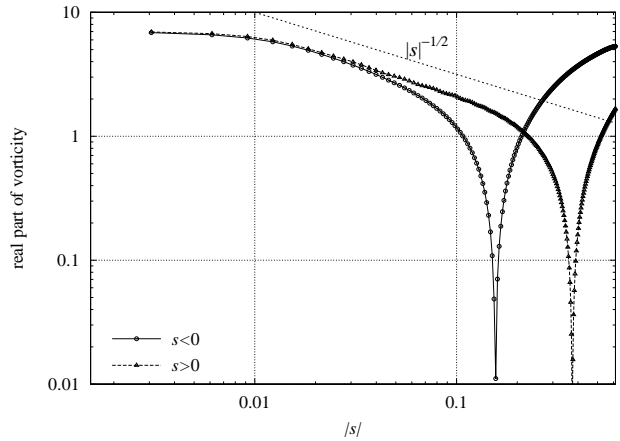,width=.48\textwidth,clip=}} 
\else\drawing 60 10 {\xfg{f:section_log} algebraic}
\fi
\caption{Real part of the vorticity along the cut shown on 
Figure~\ref{f:section_all} in log-log coordinates. A range of algebraic 
behavior can be seen.} 
\label{f:section_log}
\end{figure}

Figures \ref{f:omega_cntr} and \ref{f:omega_cntr_enlar} show the real
part of the vorticity $\omega \equiv -\nabla ^2 \psi$ at the latest
time when the  behavior in the parareal domain is still
sufficiently smooth to be unaffected by truncation, that is when the
complex singularities are not too close to the parareal domain chosen.
It is seen that the solution in the near-singular region is almost one
dimensional and hence has strongly depleted nonlinearity. The
structure of the region of high vorticity gradients suggests that the
singular set is a smooth curve. A cross section of the vorticity in the 
near-singular region is shown in Figure~\ref{f:section_all}; a blowup 
of the region of highest vorticity
(cf. Figure~\ref{f:section_log}) 
shows about one decade of an approximately one-over-square root behavior
(exponent $-1/2$) 
of the vorticity when crossing the singular manifold. This implies
some range of square-root behavior for the velocity. 

To check on errors due to truncation
and filtering we changed the resolution from $512^2$ to $1024^2$ and 
$2048^2$ modes without changing any of the physical parameters (but, of course,
adapting the time step to the spatial mesh). The results differed by less than
the filtering level, a good  indication of reliability. We also tried to
extend the scaling range at the higher resolutions by chosing a slightly
later output time, so as to let the singularities move closer to the parareal
plane. 
In fact, the scaling range did not increase appreciably, perhaps
because of roundoff problems. We cannot therefore ascertain that the 
vorticity truly diverges with
exponent $-1/2$ when approaching the singular set.

\section{Asymptotic analysis of 2-D complex singularities}
\label{s:asymptotics}

In Ref.~\cite{kuramoto} the following result was established: the inviscid
one-dimensional Burgers equation with initial conditions which are
trigonometric polynomials in the space variable has, at short real times,
square root branch point singularities for the velocity, which are within a
distance $\delta(t) \propto \ln (1/t)$ of the real domain. It was also briefly
pointed out that the $\ln (1/t)$ law can probably be extended to the 3-D Euler
equation with initial conditions which are trigonometric polynomials in the
space variables (such as the Taylor--Green flow). Further results on
complex singularities 
for the Burgers equation may be found in 
Refs.~\cite{bessisfournier84,bessisfournier90}.

We now show that the same law applies to the 2-D Euler equation at short
times and, furthermore, we give a consistency argument shedding some
light on the square root behavior  reported in the previous section. 
For simplicity we
shall limit ourselves to the two-mode initial condition \rf{deuxmodes} used in
the previous section, although most of the arguments can be generalized to
arbitrary trigonometric polynomials.

We start with the 2-D Euler equation written in stream function formulation
\begin{equation}
  \partial_t \nabla^2 \psi = J(\psi,\nabla^2 \psi),
  \label{streamform}
\end{equation}
where $J(f,g)\equiv \partial_1f\partial_2g
-\partial_1g\partial_2f$. It is easily shown 
that \rf{streamform} has a solution in the
form of a temporal Taylor series 
\begin{equation}
  \psi(\x,t) = \sum_{n\ge0} \psi_n(\x)\, t^n,
  \label{shorttimeexpansion}
\end{equation}
where $\psi_0$ is the initial condition and the $\psi_n(\x)$'s for $n\ge 1$
are defined recursively by
\begin{equation}
  \nabla^2\psi_{n+1} = \frac{1}{n+1} \sum_{m+p=n} J(\psi_m, \nabla^2\psi_p).
  \label{recurrencepsin}
\end{equation}
For the two-mode initial condition $\psi_0(\x)=\cos(x_1)+\cos(2x_2)$, it is
easily checked that $\psi_{n+1}$ is a trigonometric polynomial which, when
written in terms of complex exponentials, involves only modes of the form $\e
^{ipx_1+2iqx_2}$ where $p$ and $q$ are signed integers such that $|p|+|q|\le
n+1$.  Each term can now be continued from real $\x$ to complex $\z=\x+i\y
=(z_1,\,z_2)$.

Trigonometric polynomials are instances of entire functions. Hence, initially,
$\delta=\infty$ and, by continuity, $\delta(t)$ will be large at small real
times. Singularities will thus be present only for suitably large $|\y|$. For
large positive $y_1$ and $y_2$, the dominant contributions to $\psi_n(\x+i\y)$
have $p<0$ and $q<0$.  (For our special choice of initial conditions, symmetry
arguments make it unnecessary to examine the other three sign quadrants in the
$\y$ plane.)  When $y_1\to +\infty$ and $y_2 \to +\infty$, the dominant terms
in $\psi_n(\z)$ are of the form
\begin{equation}
  \psi_n(\z) \simeq \sum_{k=1}^n \beta^n_k \e ^{-i \left[ kz_1 +
  2(n+1-k)z_2 \right]},
\label{asymptpsi-n}
\end{equation}
where the complex coefficients $\beta^n_k$ satisfy suitable
recursion relations, not needed here. If we formally keep  only those dominant
terms in \rf{shorttimeexpansion}, we obtain 
\begin{eqnarray}
\psi(\z,t)&\simeq& \frac{1}{2}\left(\e ^{-iz_1}+ \e ^{-2iz_2} \right)
    \nonumber \\
&& +\frac{1}{t} \sum_{n\ge0} \sum_{k=1}^n \beta^n_k \left(t\e^{-iz_1}\right)^k
    \left( t \e^{-2iz_2}\right)^{n+1-k}.
\label{asymptpsi}
\end{eqnarray}
We now let $t\to 0$ and simultaneously $|\y|\to \infty$ in such a way that
$t\e ^{-iz_1}$ and $t\e ^{-2iz_2}$ stay finite and we find that (i) all the
terms in the Taylor expansion stay finite and (ii) all the terms not included
in \rf{asymptpsi} are subdominant. This observation leads us naturally to
making the asymptotic ansatz $\psi(\z,t) \simeq (1/t) F(\tz)$ where
\begin{equation}
\tz = (\tilde z_1,\, \tilde z_2) \equiv (z_1+i\ln t,\, z_2+(i/2) \ln t).
\label{deftz}
\end{equation}
Straightforward substitution into \rf{streamform} leads to
\begin{equation}
\tnabla ^2 (-1+i\tpartial_1+(i/2)\tpartial_2) F = \tJ(F, \tnabla ^2 F),
\label{asympteuler}
\end{equation}
where the overscript tilde means that the partial derivatives are taken with 
respect to the new variables. The initial condition \rf{deuxmodes} becomes now
a boundary condition
\begin{equation}
F(\tz) \simeq \frac{1}{2} \left( \e ^{-i\tilde z_1} + \e ^{-2i\tilde z_2}
\right), \quad \tilde y_1\to -\infty,\,\,\,\,\tilde y_2\to -\infty.
\label{asymbound}
\end{equation}
If \rf{asympteuler} with this boundary condition has a unique solution
possessing  complex space singularities at a finite distance from the real
domain, then it follows immediately from the change of variables
\rf{deftz} that
$\delta(t) \propto \ln (1/t)$ as $t \to 0$ in the original variables.
We have recently checked  numerically  that, for
the 2-D flow with initial condition given by (\ref{deuxmodes}), 
this scaling law holds over more than 8 decades, up to $t\approx
0.1$. Details will be published elsewhere together with numerical
solutions of the asymptotic equation \rf{asympteuler}.

The numerical results of the previous section suggest that, for fixed $t$, the
singularities in $\tcset ^2$ are located on one-dimensional complex manifolds,
near which the velocity has 
some range of square root behavior.

Using the short-time asymptotic equation \rf{asympteuler}, we show now that
the observed behavior is consistent with the two-dimensional Euler
equation. For this, assume that the (rescaled) stream function $F(\tz)$ can
be represented, at least approximately, as  

\begin{equation}
F(\tz)= G(\tz) + \Phi ^\alpha(\tz)  + {\rm h.o.t.},
\label{singexp}
\end{equation}
where $G$ and $\Phi$ are analytic functions, $\alpha$ is a non-integer exponent,
$\Phi$ (but not $G$) vanishes on the singular manifold ${\cal S}$ and
h.o.t.\ stands for ``higher order terms'', that is terms involving higher
powers of $\Phi$. The expansion \rf{singexp} is somewhat reminiscent 
of  the singular expansion used in Refs.~\cite{tanveer-speziale,BNXW}, 
except that (i) we do not a priori suppose that the singular manifold 
$\Phi=0$ moves with the flow and (ii) we shall determine the value of the 
exponent $\alpha$. Substituting \rf{singexp} into \rf{asympteuler}, we obtain
\begin{eqnarray}
&&\tnabla ^2 (-1+i\tpartial_1+\frac{i}{2}\tpartial_2) G - \tJ (G,\tnabla^2 G) =
\nonumber \\ &&  \gamma\left [ (\tpartial_1\Phi)^2 +
(\tpartial_2\Phi)^2 \right ] \left [ 
\tJ (G,\Phi)-i(\tpartial_1+\frac{1}{2}\tpartial_2)\Phi  \right ]\Phi^{\alpha - 3} \nonumber \\ && + \alpha ^2 (\alpha -1)
\tJ (\Phi,(\tpartial_1\Phi)^2+(\tpartial_2\Phi)^2) \Phi^{2\alpha -2} + 
{\rm h.o.t.},
\label{equationphiF}
\end{eqnarray}
where $\gamma \equiv\alpha(\alpha-1)(\alpha-2)$. In \rf{equationphiF} the most
singular term near ${\cal S}$ is that involving $\Phi^{\alpha - 3}$, which 
cannot be balanced by any other term. Hence, its coefficient must vanish,
thereby constraining the  functions $G$ and $\Phi$ to satisfy
$ \tJ (G,\Phi)-i(\tpartial_1+\frac{1}{2}\tpartial_2)\Phi=0$. For non-integer
$\alpha$, the term involving  $\Phi^{2\alpha -2}$ also cannot
be balanced by any other term unless it is actually analytic, that is $2\alpha
-2$ is an integer. The smallest possible value 
is $\alpha =3/2$. The velocity has then exponent $1/2$,
that is a square root behavior near ${\cal S}$.  
This is here derived under the assumption that $\Phi$ vanishes
linearly near the singular manifold. Otherwise a different scaling
law may be obtained.

\section{Conclusion}
\label{s:conclusion}

It is clear that the issue of finite-time blowup is still open for
initially smooth 3-D Euler flow (and also for 3-D Navier--Stokes
flow).  We have here proposed investigating this issue within the more
restricted class of initially analytic flow, for which (hypothetical)
real singularities are necessarily preceded by singularities in
complex space.  For this purpose we propose a new spectral adaptive
strategy which requires the tracking of complex singularities and the
use, at suitable times, of a regularizing map sending singularities
too close to the real domain away from it.

Finally, we should mention that the issue of experimental
study of 3-D Euler blowup was discussed at the workshop. We 
noted the following: if a flow is started by 
standard methods such as a moving grid, the ``initial conditions''
will have considerable small-scale excitation (viscous boundary layers 
generated at solid boundaries) and would in fact become singular if
this was extrapolated to zero viscosity. Alternative ways may
be tried where the flow is put in motion initially by body forces with
no small-scale component, such as electromagnetic or (ultra)sonic stirring.

\vspace*{1mm}
\par\noindent {\bf Acknowledgments}

We are grateful to Claude Bardos and Peter Constantin for very useful
discussions and remarks. Special thanks are due to Richard Pelz for many
discussions and references; his recent untimely death is a great loss
to us. Many thanks are also due to Robert Kerr for providing us with 
a number of  references on blowup.
Computational resources were provided by the Yukawa Institute (Kyoto).
This research was supported by the Department of Energy, under
contract W-7405-ENG-36, by the European Union
under contract HPRN-CT-2000-00162, by the Indo-French Centre for the
Promotion of Advanced Research (IFCPAR~2404-2),  by the Japan
Scholarship Foundation and by the Turbulence Working Group (Los Alamos).

\end{document}